\documentclass[twocolumn,showpacs,preprintnumbers,amsmath,amssymb,superscriptaddress,balancelastpage,prl]{revtex4}

\include{options}

\begin{document}

\title{General analysis of the ARPES line shape for strongly correlated electron systems}
\author{S.~G.~Ovchinnikov}
\email{sgo@iph.krasn.ru}
\author{E.~I.~Shneyder}
\affiliation{Kirensky Institute of Physics SB RAS, 660036 Krasnoyarsk, Russia}
\author{A.~A.~Kordyuk}
\affiliation{Institute of Metal Physics of National Academy of Sciences of Ukraine, 03142 Kyiv, Ukraine}
\date{\today}


\begin{abstract}

In many cases the standard perturbation approach appears to be too simple to describe precisely the angle resolved photoemission spectrum of strongly correlated electron system. In particular, to describe the momentum asymmetry observed in photoemission spectra of high-$T_c$ cuprates a phenomenological approach based on extremely correlated Fermi-liquid model has been recently introduced. 
Here we analyze the general structure of the Green function of quasiparticles in strongly correlated electron systems  and stress that it is defined not only by the self-energy of Hubbard quasiparticles but also by a strength operator. The later leads to an additional odd momentum contribution to the spectral function and alone can explain the observed asymmetry. So, the asymmetry of the ARPES spectra can be a measure of the strength of electron correlations in materials.

\end{abstract}


\pacs{71.10.Ay, 74.25.Jb, 74.72.Gh}
\maketitle

{\it Introduction.---}Angle-resolved photoemission spectroscopy (ARPES) is very powerful method to study the electronic structure of solids, especially for anisotropic layered materials like high-Tc cuprates where the quasiparticle (QP) dispersion law $ \epsilon \left( {{k_x},{k_y}} \right) $ may be obtained \cite{ARPES03}. It is particularly important for strongly correlated electron systems (SCESs) where conventional {\it ab initio} local density approximation to the density functional theory (LDA-DFT) fails to get the correct QP electronic structure.

Within the usually applied three step model of photoemission and using the sudden approximation \cite{SolStatPhys1969}, the photoelectron counts $ I\left( {{{\vec k}},\omega } \right) $ as a function of energy $ \omega $ and momentum $ {\vec k} $  are given by $I\left( {\vec k,\omega } \right) = {\left| {{M_{ij}}} \right|^2}f\left( \omega  \right)A\left( {\vec k,\omega } \right)$, where $ {M_{ij}} $ is the dipole matrix element for the photo-excitation, $ f\left( \omega  \right) $ is the Fermi-Dirac distribution, and $ A\left( {{{\vec k}},\omega } \right) = \left( { - {1 \mathord{\left/ {\vphantom {1 \pi }} \right.
\kern-\nulldelimiterspace} \pi }} \right){\mathop{\rm Im}\nolimits} G\left( {{{\vec k}},\omega } \right)$  is the spectral function for the single-electron retarded Green function $ G\left( {{\vec k},\omega } \right) $. If one disregards the effect of the energy and momentum resolutions as well as the matrix elements effect \cite{Borisenko2001} and the extrinsic background \cite{Kaminski2004}, the photoelectron intensity is proportional to the spectral function  multiplied by the Fermi function.

Usually the standard perturbation representation of the Green function $ G\left( {{\vec k},\omega } \right) $ in terms of Fermi-type operators is used for the ARPES analysis \cite{ARPES03,Kordyuk06}. Introducing the real and imaginary parts of the QP self-energy $ {\Sigma _{\vec k}}\left( \omega  \right) = {\Sigma _{\vec k}}^{'}\left( \omega  \right) + i{\Sigma _{\vec k}}^{''}\left( \omega  \right) $ one can write down the spectral function
\begin{equation}
\label{SpectrFunc}
A\left( {\vec k,\omega } \right) =  - \frac{1}{\pi }\frac{{{\Sigma _{\vec k}}^{''}\left( \omega  \right)}}{{{{\left[ {\omega  - {\varepsilon_{\vec k}} - {\Sigma _{\vec k}}^{'}\left( \omega  \right)} \right]}^2} + {\Sigma _{\vec k}}^{''}{{\left( \omega  \right)}^2}}}.
\end{equation}
This is the central formula for the ARPES analysis. It provides the Lorentzian line shape for the momentum distribution curve (MDC) defined as $ I\left( {{\vec k},\omega  = {const}} \right) $ as long as the self-energy $ \Sigma _{\vec k}\left( \omega  \right) $ can be considered as momentum independent and the bare dispersion $\varepsilon_{\vec k}$ is linearized in the vicinity to the Fermi level \cite{VallaS1999}. (More general assumption \cite{Randeria2004} is that $\frac{\partial {\Sigma _{\vec k}}^{'}\left( \omega  \right)}{\partial k}$ can only be a constant independent of $\omega$.)

This approach works well for metals where Fermi-liquid picture is adequate and for many high-$T_c$ cuprates \cite{ARPES2005,ARPES03,Kordyuk06}.
Nevertheless from theoretical point of view the standard perturbation approach seems to be non appropriate for SCESs such as underdoped and optimally doped hole cuprates where non Fermi liquid effects have been found in the pseudogap state. Various approaches towards clarifying the QP properties in the regime of strong electron correlations have been attempted: a phenomenological marginal Fermi liquid approach \cite{Varma1989}, an asymptotic solution to the Gutzwiller projected ground state of the $ t-J $ model \cite{Anderson08}, low dimensional non-Fermi-liquid theory \cite{OrgadKivelson}. Recently the extremely correlated Fermi-liquid model has been suggested \cite{Shastry11}. Its improved phenomenological version \cite{MatsuyamaGweon} successfully describes the dichotomy of the spectral functions of momentum and energy. One important result obtained in paper \cite{MatsuyamaGweon} is the
MDC asymmetry that has been observed for under- and optimally doped La$_{2-x}$Sr${_x}$CuO$_4$ \cite{Yoshida07} and Ca$_{2-x}$Na$_x$CuO$_2$Cl$_2$ \cite{KyleShen2005}.

In this letter we have shown that the MDC asymmetry is not a particular model \cite{Shastry11,MatsuyamaGweon} property. It is a general property of the spectral function in SCESs, where the Coulomb interaction $ U $ is much larger then QP kinetic energy and the perturbation over $ {{{\varepsilon_{\vec k}}} \mathord{\left/
 {\vphantom {{{\varepsilon_{\vec k}}} {U \ll 1}}} \right.
 \kern-\nulldelimiterspace} {U \ll 1}} $
seems to be more appropriate.

{\it Formalism.---}The natural and proper mathematical tool in the atomic limit $ {{{\varepsilon_{\vec k}}} \mathord{\left/
 {\vphantom {{{\varepsilon_{\vec k}}} {U \ll 1}}} \right.
 \kern-\nulldelimiterspace} {U \ll 1}} $ is given by the Hubbard X-operators \cite{Hubbard65}. Their algebra automatically fulfills the constraint condition that forbids some sectors of the Hilbert space due to strong electron correlations. Formerly the Hubbard's ideas of $ X $-operators were developed in cluster perturbation theory within the generalized tight-binding (GTB) method \cite{GTB1989,SGO1994}. The GTB approach has been proposed to calculate the electronic structure of correlated materials like underdoped cuprates, manganites,  and cobaltites \cite{RevGTB}. It's  {\it ab initio} LDA+GTB version \cite{GTBplusLDA} is a hybrid scheme used the local density approximation to construct the Wannier functions and obtain the single electron and Coulomb parameters of the multiband Hubbard-like Hamiltonian. At the next step this method combines the exact diagonalization of the intracell part of the Hamiltonian, construction of the Hubbard operators on the basis of the exact intracell multielectron eigenstates, and the perturbation treatment of the intercell hoppings and interactions.

This is essentially multielectron approach which does not use the idea of bare electron. An electron in GTB is a linear combination of QP excitations between multielectron initial $ {d^n} $ - and final  $ {d^{n \pm 1}} $ configurations. Each excitation from initial state $ \left| q \right\rangle $ to final state  $ \left| p \right\rangle $ is described by the Hubbard operator $ X_f^{pq} = \left| p \right\rangle \left\langle q \right| $. Thereby any local operator can be represented as a linear combination of $ X $-operators. So the operator of removing of electron with spin $ \sigma $ at a lattice site $ f $  takes the form
\begin{eqnarray}
\label{aXRelation}
c_{f,\sigma }& =& \sum\limits_{p,q} {\left| p \right\rangle \left\langle p \right|} {c_{f,\sigma }}\left| q \right\rangle \left\langle q \right| = \nonumber \\
 &=&\sum\limits_{p,q} {{\gamma _{\sigma }}\left( {p,q} \right)X_f^{pq}}=\sum\limits_m {{\gamma _{\sigma }}\left( m \right)X_f^m}.
\end{eqnarray}
To simplify notations we introduce the QP band index $ m $ corresponding to the pair $ \left( {p,q} \right) $. Equation (\ref{aXRelation}) clearly shows the difference between Fermi type quasiparticle description in the single electron language and the multielectron one.
The operator $ {c_{f,\sigma }} $ decreases the number of electrons by one for all sectors of the Hilbert space simultaneously, while the $ {X_f^{m}} $ operator describes the partial process of electron removing from the $ \left( {N} \right) $-electron configuration $ \left| q \right\rangle $, with the final $ \left( {N-1} \right) $-electron configuration $ \left| p \right\rangle $.
The matrix element $ \gamma _{\sigma }\left( m \right) $ gives the probability of such a process. It should be noted that splitting of an electron onto different Hubbard fermions stated by Eq. (\ref{aXRelation}) and the following spectral weight redistribution over these quasiparticles are the underlying effects of the band structure formation in correlated systems.

According to Eq. (\ref{aXRelation}) the single electron retarded Green function $ {G}\left( {\vec k,\omega } \right)={\left\langle {{\left\langle {a_{\vec k,\sigma}} \right.}}
 \mathrel{\left | {\vphantom {{\left\langle {a_{\vec k,\sigma}} \right.} {\left. {\mathop {a_{\vec k,\sigma}}\limits^\dag  } \right\rangle }}}
 \right. \kern-\nulldelimiterspace}
 {{\left. {\mathop {a_{\vec k,\sigma}}\limits^\dag  } \right\rangle }} \right\rangle _\omega } $ is given by a linear combination of the Green functions of Hubbard quasiparticles $ D_{\vec k,\omega }^{mn} = {\left\langle {{\left\langle {X_{\vec k}^m} \right.}}
 \mathrel{\left | {\vphantom {{\left\langle {X_{\vec k}^m} \right.} {\left. {\mathop {X_{\vec k}^n}\limits^\dag  } \right\rangle }}}
 \right. \kern-\nulldelimiterspace}
 {{\left. {\mathop {X_{\vec k}^n}\limits^\dag  } \right\rangle }} \right\rangle _\omega } $,
\begin{eqnarray}
\label{CreenFunc}
G\left( {\vec k,\omega } \right) = \sum\limits_{m,n} {{\gamma _\sigma }\left( m \right)\gamma _\sigma ^*\left( n \right)} D_{\vec k,\omega }^{mn},
\end{eqnarray}
here the notation of Zubarev \cite{Zubarev1960} for Green functions is used.
Due to a complicate commutation rules there is no conventional Wick’s theorem and conventional diagram technique for Hubbard operators. Nevertheless the generalized Wick’s theorem has been proved \cite{WickTheorem1973} and then diagram technique for $ X $-operators developed \cite{Zaitsev1975,IzyumovLetfullov}.
The Dyson equation \cite{SGOandVVV} for the matrix Green function $ \hat D\left( {\vec k,\omega } \right) = \left\{ {D_{\vec k,\omega }^{mn}} \right\} $ has also been modified
\begin{subequations}\label{DysonXeq}
\begin{align} \label{DysonXeqA}
\hat D\left( {\vec k,\omega } \right) = { {  \mathcal {\hat G}}_{\vec k}\left( \omega  \right) }{\hat P_{\vec k}}\left( \omega  \right),
\end{align}
with propagator ${  \mathcal {\hat G}}_{\vec k}\left( \omega  \right)$ is
\begin{align} \label{DysonXeqB}
{  \mathcal {\hat G}}_{\vec k}\left( \omega  \right) = {\left[ {{  \mathcal {\hat G}}_0^{ - 1}\left( \omega  \right) - {{\hat P}_{\vec k}}\left( \omega  \right){{\hat t}_{\vec k}} - {{\hat \Sigma }_{\vec k}}\left( \omega  \right)} \right]^{ - 1}}.
\end{align}
\end{subequations}
Here $ {  \mathcal {\hat G}}_{0}^{ - 1}\left( \omega  \right) $ is a local propagator determined by the multielectron eigenstates  $ \left| p \right\rangle $ and $ \left| q \right\rangle $, $ {{{\hat t}_{\vec k}}} $ is interaction matrix with elements $ t_{\vec k}^{mn} = {\gamma _\sigma }\left( m \right)\gamma _\sigma ^*\left( n \right){\varepsilon_{\vec k}} $, where $ {\varepsilon_{\vec k}} $ is the bare band dispersion.
It should be stressed that function $ {\hat \Sigma _{\vec k}}\left( \omega  \right) $ in Eq. (\ref{DysonXeqB}) is the self-energy for the Hubbard fermions and therefore it is different from the single-electron one in Eq. (\ref{SpectrFunc}).

Besides the self-energy $ {\hat \Sigma _{\vec k}}\left( \omega  \right) $ of Hubbard quasiparticles the unusual strength operator $ {\hat P_{\vec k}}\left( \omega  \right) $ appears in Eq. (\ref{DysonXeq}). It results both in the redistribution of the QP spectral weight and in renormalization of QP dispersion which becomes dependent on doping and temperature. Initially strength operator has been introduced in the diagram technique for spin operators \cite{Baryakhtar}. It is important that in order to use the generalized Dyson equation in the perturbation expansion it is necessary to calculate both functions $ {\hat \Sigma _{\vec k}}\left( \omega  \right) $ and $ {\hat P_{\vec k}}\left( \omega  \right) $ in the same order of perturbation \cite{VVVDDM2008}.

{\it Spectral function.---}The dimension of $ \hat D\left( {\vec k,\omega } \right) $ matrix depends on the energy interval under consideration. For example in cuprates only one kind of QPs is involved in the low excitation energy limit of ARPES. To demonstrate that general structure of QP Green functions in perturbation theory for SCESs results in additional odd contribution to the spectral function $A\left( {\vec k,\omega } \right)$ we proceed with above case. This involves no loss of generality. For hole-doped cuprates $ m = \left\{ {\left( { - \sigma ,2} \right)} \right\} $, where doublet $ \left| \sigma  \right\rangle $ and singlet $ \left| 2  \right\rangle $ are the ground terms of the CuO$_4$ unit cell with $1$ and $2$ holes per site, respectively.
In this low energy limit the exact single electron Green function reads
\begin{eqnarray}
\label{CuprGrFunc}
G\left(\vec k,\omega \right) = {\left| {{\gamma _{\bar \sigma ,2}}} \right|^2}\frac{{{P_{\vec k,\omega }}}}{{\omega  - \varepsilon  - {t_{\vec k}}{P_{\vec k,\omega }} - {\Sigma _{\vec k,\omega }}}},
\end{eqnarray}
where $\varepsilon  = {\varepsilon _0} - \mu $ is eigenvalue of local state $ \left| \sigma  \right\rangle $ and $ \mu $ is chemical potential. Generally both self-energy and strength operator can be presented as a sum of real $ P_{\vec k, \omega}^{'} $, $ \Sigma_{\vec k, \omega}^{'} $ and imaginary $ P_{\vec k, \omega}^{''} $, $ \Sigma_{\vec k, \omega}^{''} $ parts, respectively. Therefore electron spectral function takes the form
\begin{eqnarray}
\label{TotalSpectr}
A\left(\vec k,\omega \right) = {\frac{{\left| {{\gamma _{\bar \sigma 2}}} \right|}}{\pi }^2} \times \nonumber \\
 \left( {\frac{{P_{\vec k,\omega }^{'}{\Gamma _{\vec k, \omega}}}}{{{{\left(\omega  - {\epsilon _{\vec k, \omega}}\right)}^2} + \Gamma _{\vec k, \omega}^2}} + \frac{{{\rm P}_{\vec k,\omega }^{''}\left(\omega  - {\epsilon _{\vec k, \omega}}\right)}}{{{{\left(\omega  - {\epsilon _{\vec k, \omega}}\right)}^2} + \Gamma _{\vec k, \omega}^2}}} \right),
\end{eqnarray}
here $ {\epsilon _{\vec k, \omega}} = \varepsilon + {t_{\vec k}}P_{\vec k,\omega }^{'} + \Sigma _{\vec k,\omega }^{'} $ is renormalized QP dispersion and $ {\Gamma _{\vec k, \omega}} = {t_{\vec k}}P_{\vec k,\omega }^{''} + \Sigma _{\vec k,\omega }^{''} $ is inverse life time of QP's.

Being interested to analyze the MDC line shape determined by Eq. (\ref{TotalSpectr})  we fix energy $ \omega=\omega_0 $ and assume $k$ independence of QP inverse life time $ \Gamma _{\vec k, \omega_0} $ and strength operator $ P _{\vec k,\omega_0 } $ in a small vicinity to the Fermi level.
In such a case the spectral function $ A\left(\vec k,\omega_0 \right) $ appears to be a sum of even $ A^{evn}\left(\vec k,\omega_0 \right) $ and odd $ A^{odd}\left(\vec k,\omega_0 \right) $ contributions.
In the limit $ {\Gamma _{\omega_0}} \to 0 $ the even part tends to a  $ \delta  $-function with renormalized spectral weight $ {\left| {{\gamma _{\bar \sigma 2}}} \right|^2}P_{\omega_0 }^{'} $. For the finite QP inverse life time $ {\Gamma _{\omega_0}} $ and linearized QP dispertion $ \tilde{\epsilon} _{\vec k, \omega_0} $ the even part has the Lorentzian line shape similarly to the case of non correlated Fermi liquid
\begin{eqnarray}
\label{EvenSpectr}
A^{evn}\left(\vec k,\omega_0 \right) = {\frac{{\left| {{\gamma _{\bar \sigma 2}}} \right|}}{\pi }^2} \cdot
 {\frac{{P_{\omega_0 }^{'}{\Gamma _{\omega_0}}}}{{{{\left(\omega_0 - {\upsilon _{F}\left( k-k_{F} \right)} \right)}^2} + \Gamma _{\omega_0}^2}}}.
\end{eqnarray}
However the most peculiar feature of the spectral function in SCEC's is the odd contribution that appears in Eq. (\ref{TotalSpectr}) due to imaginary part $ P_{\vec k,\omega }^{''} $ of the strength operator,
\begin{eqnarray}
\label{OddSpectr}
A^{odd}\left(\vec k,\omega_0 \right) = {\frac{{\left| {{\gamma _{\bar \sigma 2}}} \right|}}{\pi }^2} \cdot
  { \frac{{{\rm P}_{\omega_0 }^{''}\left(\omega  - {\tilde{\epsilon} _{\vec k, \omega_0}}\right)}}{{{{\left(\omega  - {\tilde{\epsilon} _{\vec k, \omega_0}}\right)}^2} + \Gamma _{ \omega_0}^2}}}.
\end{eqnarray}
The strength operator results in Eq. (\ref{DysonXeq}) from non-Fermi commutation rules of the Hubbard $X$-operators as well as for spin Green function it results from non-Bose commutation rules of the spin operators \cite{Baryakhtar}. In the limit of weak correlations anticommutator (commutator) of Fermi (Bose)-like operators of Hubbard's quasiparticles is  equal to c-number. Formally in this limit strength operator tends to unit $ {{\rm P}_{\vec k,\omega }} \to 1 $ and odd contribution to the spectral function disappears.

To shortly discuss the problem in the superconducting state we write down the matrix Green function $ {\hat D \left( \vec k,\omega \right) } = {\left\langle {\left\langle {{\Psi _{\vec k\sigma }}} \right|\left. {\Psi _{\vec k\sigma }^\dag } \right\rangle } \right\rangle _\omega } $ in terms of the Nambu operators $ \Psi _{\vec k\sigma }^\dag  = \left( {X_{\vec k}^{\sigma 0},X_{ - \vec k}^{0, - \sigma }} \right) $ and then denote components of all relevant matrices via the corresponding superscript.
According to Eq.  (\ref{DysonXeqA}) the normal state function $ D_{ \vec k,\omega }^{\left( {11} \right)} = {\left\langle {\left\langle {X_{\bf{k}}^{0\sigma }} \right|\left. {X_{\bf{k}}^{\sigma 0}} \right\rangle } \right\rangle _\omega } $ is given by expression
\begin{subequations}\label{D11}
\begin{align} \label{D11_tot}
D_{ \vec k,\omega }^{\left( {11} \right)} = {\mathcal {G}}_{ \vec k,\omega }^{\left( {11} \right)} P_{ \vec k,\omega }^{\left( {11} \right)} + {\mathcal {G}}_{ \vec k,\omega }^{\left( {12} \right)} P_{ \vec k,\omega }^{\left( {21} \right)},
\end{align}
where propagator functions $ {\mathcal {G}}_{ \vec k,\omega }^{\left( {11} \right)} $ and $ {\mathcal {G}}_{ \vec k,\omega }^{\left( {12} \right)} $ are solutions of the Eq. (\ref{DysonXeqB})
\begin{align}
\label{G11}
{\mathcal { G}}_{\vec k,\omega }^{\left( {11} \right)} =  \frac{1}{\det {\mathcal {\hat G}}_{ \vec k,\omega } }
 { \left( \omega  + \varepsilon  - t_{\vec k}^{ \left( {22} \right) }P _{\vec k,\omega }^{\left( {22} \right)} - \Sigma _{\vec k,\omega }^{\left( {22} \right) } \right) }, \\
\label{G12}
{\mathcal { G}}_{\vec k,\omega }^{\left( {12} \right)} =  \frac{1}{\det {\mathcal {\hat G}}_{ \vec k,\omega } }
{ \left(  t_{\vec k}^{\left( {11} \right) }P _{\vec k,\omega }^{\left( {21} \right)} +\Sigma _{\vec k,\omega }^{\left( {12} \right)} \right)  },
\end{align}
and determinator reads
\begin{align}
\label{det}
\det {\mathcal {\hat G}}_{ \vec k,\omega }=  \left( {\omega  + \varepsilon - t_{\vec k}^{ \left( {22} \right) }P _{\vec k,\omega }^{\left( {22} \right)} - \Sigma _{\vec k,\omega }^{\left( {22} \right)}} \right) \times \nonumber \\
\left( {\omega  - \varepsilon - t_{\vec k}^{ \left( {11} \right) }P _{\vec k,\omega }^{\left( {11} \right)} -  \Sigma _{\vec k,\omega }^{\left( {11} \right)}} \right) + \nonumber \\
\left( \Sigma _{\vec k,\omega }^{\left( {12} \right)} + t_{\vec k}^{ \left( {22} \right) }P _{\vec k,\omega }^{\left( {12} \right)} \right) \left( \Sigma _{\vec k,\omega }^{\left( {21} \right)} + t_{\vec k}^{ \left( {11} \right) }P _{\vec k,\omega }^{\left( {21} \right)} \right).
\end{align}
\end{subequations}
In the superconducting state the off-diagonal components of the strength operator just like off-diagonal self-energy components differ from zero \cite{VVVDDM2008,VVVGolovnya}. That is way the normal state function $ D_{ \vec k,\omega }^{\left( {11} \right)}$ has so complicated structure.
We do not give the cumbersome expression for the spectral function
$ A^{\left({11}\right)}\left( {{{\vec k}},\omega } \right) = \left( { - {1 \mathord{\left/ {\vphantom {1 \pi }} \right.
\kern-\nulldelimiterspace} \pi }} \right){\mathop{\rm Im}\nolimits} D_{ \vec k,\omega }^{\left( {11} \right)}$ in the superconducting state since general analysis of its symmetry is beyond the scope of this paper. Nevertheless it is easy to show that spectral function $ A^{\left({11}\right)}\left( {{{\vec k}},\omega } \right)$ of superconductors with $d_{x^2-y^2}$-gap symmetry has in the nodal direction $k_x=k_y$ the same structure as the function given by Eq. (\ref{TotalSpectr}), to wit, the additional odd contribution to the ARPES line shape should be present.

{\it Discussions.---}
The odd contribution to the spectral functions obtained above in Eq. (\ref{TotalSpectr}) results from the general structure of the Green function of quasiparticles in strongly correlated systems. We compare it to the structure of the spectral function designed in paper \cite{MatsuyamaGweon} to describe the normal state ARPES line shape of high-T$_c$ superconductors.
This phenomenological approach based on the modified theory of extremely correlated Fermi liqued \cite{Shastry11} successfully reproduces peculiarities of ARPES line shape for different materials even such as the MDC asymmetry.
It turns out that phenomenologically found spectral function has the structure which correlates with the structure in Eq. (\ref{TotalSpectr}) based on the general background.
Namely, this spectral function \cite{MatsuyamaGweon} consists of the symmetrical contribution with the real part of strength operator defined in Hubbard-I approximation $ P_{\vec k, \omega}^{'} = 1- {\frac n 2}$, where n is the number of
electrons (holes) per unit cell,  and asymmetrical contribution which has the same structure of expression as given in Eq. (\ref{OddSpectr}) but implies some complicated expression for the imaginary part of strength operator $ P_{\vec k, \omega}^{''}$.
Examples of expressions for strength operator obtained beyond the mien field approximation can be found elsewhere \cite{VVVDDM2008} since we do not discuss experimental data.
We argue that asymmetrical structure of spectral function in strongly correlated electron systems has among other things the fundamental reason considered above and can reflect the strength of correlations.

{\it Acknowledgment.---}  SGO and EIS are thankful to Russian Science Foundation (project No. 14-12-00061), AAK is thankful to NAS of Ukraine (project 73-02-14) for financial support.


\end{document}